\documentclass[useAMS]{mn2e}
\usepackage{psfig}

\usepackage{times}

\def\bb{black--body~}
\def\bbp{black--body}

\def\spose#1{\hbox to 0pt{#1\hss}}
\newcommand\lsim{\mathrel{\spose{\lower 3pt\hbox{$\mathchar"218$}}
     \raise 2.0pt\hbox{$\mathchar"13C$}}}
\newcommand\gsim{\mathrel{\spose{\lower 3pt\hbox{$\mathchar"218$}}
     \raise 2.0pt\hbox{$\mathchar"13E$}}}

\title[Puzzled by GRB 060218]
{Puzzled by GRB 060218}

\author[Ghisellini, Ghirlanda \& Tavecchio]
{G. Ghisellini$^1$\thanks{E--mail: gabriele.ghisellini@brera.inaf.it},
G. Ghirlanda$^1$ and F. Tavecchio$^1$\\
$^{1}$Osservatorio Astronomico di Brera, via E.Bianchi 46, I-23807
Merate, Italy
}

\begin{document}

\pagerange{\pageref{firstpage}--\pageref{lastpage}} \pubyear{2002}

\maketitle

\label{firstpage}

\begin{abstract}
We study the optical--UV/X--ray spectral energy distribution
of GRB 060218 during the
prompt phase and during what seems to be the afterglow phase.
The results are puzzling, since if the opt--UV and the X--ray 
emission belong to a single \bbp, then its luminosity 
is too large, and this \bb cannot be interpreted as the
signature of the shock breakout of the supernova.
There are also serious problems in associating the
emission expected by the supernova shock breakout with
either the opt--UV or the X--ray emission.
In the former case we derive too small ejecta velocities; 
in the latter case, on the contrary, the required velocity
is too large, corresponding to the large radius of a \bb 
required to peak close to the UV band.
We then present what we think is the most conservative
alternative explanation, namely a synchrotron spectrum,
self--absorbed in the opt--UV and extending up to the
X--ray band, where we observe the emission of the most energetic
electrons, which are responsible for the exponential roll--over
of the spectrum.
The obtained fit can explain the
entire spectrum except the \bb observed in the X--rays,
which must be a separate component.
The puzzling feature of this interpretation is that the
same model is required to explain the spectrum also
at later times, up to $10^5$ s, because the opt--UV
emission remains constant in shape and also (approximately)
in normalisation. In this case the observed X--ray flux is
produced by self--Compton emission.
Thus the prompt emission phase should last for 
$\sim 10^5$ s or more.
Finally, 
we show that the \bb observed in X--rays,
up to 7000 seconds, can be  photospheric emission from the
cocoon or stellar material, energized by the GRB jet
at radii comparable to the stellar radius 
(i.e. $10^{10}$--$10^{11}$ cm), not very
far from where this material becomes transparent (e.g. $10^{12}$ cm).
\end{abstract}

\begin{keywords}
gamma rays: bursts; radiation mechanisms: thermal, non--thermal.
\end{keywords}

\section{Introduction}

The burst exploded February 18, 2006
is a low redshift burst ($z=0.033$, Mirabal et al. 2004), 
associated to the Type Ibc supernova SN2006aj (e.g. Modjaz et
al. 2006, Mazzali et al. 2006).
Due to its long duration (more than 3000 s) GRB 060218 could be
followed simultaneously with the BAT, XRT and UVOT instruments onboard
the SWIFT satellite (Gehrels et al. 2004).  
The 0.3--10 keV X--ray light curve
(followed by XRT starting 157 s after the BAT trigger) presents three
main phases (Campana et al. 2006, hereafter C06): phase X1: a smooth/long
($\sim 3\times 10^3$ s) bump peaking at $\sim 10^3$ s whose
time--integrated spectrum shows a non--thermal component (power-law
with an exponential cutoff) peaking (in $\nu F_\nu$) in the X-ray
band ($E_{\rm peak}\sim 5$ keV) and a \bb which comprises
about 20\% of the total 0.3--10 keV flux and dominates the soft X-ray
energy band ($kT\sim 0.13$ keV); X2: a steep power--law (or
exponential) time--decay up to 10$^4$ s still showing a slightly
softer ($kT\sim 0.1$ keV) \bb component (comprising
about 50\% of the total flux) together with a softer non--thermal
component; X3: a shallower flux decay ($\propto t^{-1.2}$) starting
at 10$^4$ s, and lasting up to several days, with a very soft (energy
spectral index $\alpha\sim 2.3$; Cusumano et al. 2006)
non--thermal component.  
The optical--UV light curve presents 2 phases: in phase UV1
there is a slow increase of the
flux, peaking at $\sim 8\times10^4$ s, followed by a fast decay up to
$\sim 1.5\times 10^5$ s; in phase UV2 there is a second bump peaking at 
$\sim$10 days
showing the typical spectral signatures of the underlying supernova
(Ferrero et al. 2006; Mirabal et al. 2006; Modjaz et al. 2006;
Sollerman et al. 2006) and suggesting photospheric expansion
velocities of 2$\times 10^4$ km s$^{-1}$ (e.g. Pian et al. 2006). 
Finally, in the radio band, the flux 
between 2 and 22 days shows a typical power law decay 
($\propto t^{-0.8}$, Soderberg et al. 2006).

Despite the available wealth of information, some of the 
observed properties of GRB 060218 are not yet understood.
In fact, although the radio flux could be due 
to external shocks, the radio spectrum at 5 days is 
inconsistent with the strong X--ray emission at the same 
epoch (Soderberg et al. 2006).
Moreover, the typical late time X--ray light curve decay (phase X3) 
is hardly reconcilable with the extremely soft spectrum in
the framework of the external shock model for GRB afterglows. 
This suggests that the late time X--ray emission might be produced 
by a continued activity of the central engine 
(i.e. ``central engine afterglow'' -- Fan, Piran \& Xu 2006).

\section{The spectral energy distribution of GRB 060218}


The most striking characteristic of GRB 060218 is perhaps the 
observation, in
the XRT 0.2--10 keV band, of a quasi--steady \bb component at a
temperature of $\sim$0.18--0.1 keV, observed up to 7000 s
(i.e. phases X1 and X2) and with a total energy of $\sim 10^{49}$ erg. 
At opt--UV frequency as well, the emission is well described by
the Rayleigh--Jeans tail of a \bb spectrum up to $10^5$ s after
trigger (phase UV1). 
It has been proposed (C06) that the
opt--UV (UV1) {\it and} the X--ray (X1 and X2) emission
are produced by the same process: the shock breakout of the supernova.

In Fig. 1 we show the optical
\footnote{
The opt--UV data shown in C06 are not de--absorbed, 
and are in the form of specific fluxes multiplied
by the FWHM widths of the different UVOT filters 
[i.e. what is plotted is $F=F(\lambda)\Delta \lambda$].
To convert in $\nu F_\nu$ fluxes, we have used
$\nu F_\nu=\lambda F(\lambda)$.
} 
to X--ray SED of GRB 060218 roughly corresponding to the same 
three epochs (X1, X2 and X3) described in the introduction.
Both the UVOT and the XRT data have been de--absorbed, 
using the same $N_{\rm H}$ and $E(B-V)$ values given in C06.
However, one can see that the optical--UV data, independently
of the assumed extinction, are above the extrapolation of
the \bb emission observed in XRT.
Furthermore, the opt-UV data de--reddened using the values in C06
describe, at early times, a Rayleigh--Jeans tail of a \bbp.

If the X--ray emission and the opt--UV flux belong to the same \bb
component, then the derived \bb luminosity is huge, exceeding
$10^{48}$ erg s$^{-1}$.   
This luminosity, if produced by the subrelativistic SN shock breakout,
should not be boosted by relativistic effects.
Since this luminosity would last for $\sim 10^4$ s, we would then 
infer a total radiated energy exceeding $10^{52}$ erg.  
This energy is close to (or above) the total kinetic energy of the 
supernova ejecta $E_{\rm K,SN}\sim 2\times10^{51}$ erg (Mazzali et al. 2006).  
Furthermore, trying to model the X--ray data with a cut--off power 
law plus a \bb equal to the one joining the opt--UV and the X--rays 
(long-dashed lines in Fig. 1) produces an unacceptable fit.  
We therefore consider this possibility as highly unlikely.

Consider now the case of a multi--color \bbp, joining the opt--UV and
X--ray band.  
In this case the opt--UV emission, belonging to the Rayleigh--Jeans
tail of the coldest \bbp, is characterised by a very large radius (of
order of $10^{15}$ cm).  
This radius cannot correspond to the radius
at which the supernova ejecta have arrived after 2000 seconds from the
trigger, if we maintain the hypothesis that the supernova and the GRB
exploded nearly simultaneously (Mirabal et al. 2006).  

The total energetics of a \bb joining the opt--UV and X--ray data
is so large to be problematic even if it is beamed radiation produced
by a relativistically moving cocoon, since it would exceed by orders
of magnitude the energetics produced by the jet which is supposed to
energize it, and whose emission should be observed, if the bursts is
not misaligned.

\begin{figure}
\vskip -0.5 true cm
\hskip -0.7 true cm \psfig{file=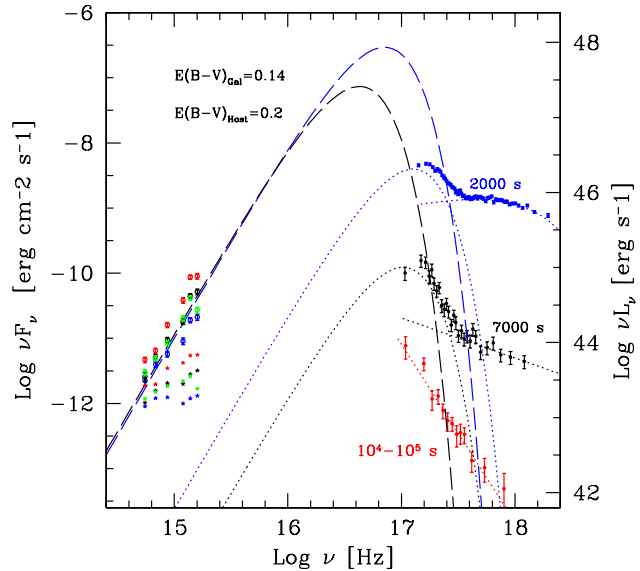,height=9cm,width=9cm}
\vskip -0.5 true cm
\caption{ 
The SED of GRB 060218 at different times.
Blue: 2000 s (integrated for $\sim 400$ s for the X--ray); 
black: 7000 s (integrated for $\sim$ 2500 s); 
red: 40,000 s; 
green: $1.2\times 10^5$ s (only UVOT data are shown).  
The opt--UV data are taken from C06 while the X--ray data have 
been re-analysed by us.
The optical--UV data lie above the blackbody found by fitting the
X--ray data (dotted lines). 
Instead, the opt--UV data seem to identify another \bb
component (long-dashed lines) which is inconsistent with the 
X--ray data at the same epochs.  
Small crosses without error bars are UVOT data not de--absorbed.
De--absorbed data [with a galactic $E(B-V)=0.14$ plus a host
$E(B-V)=0.2$] are shown with error bars. 
}
\label{fig1}
\end{figure}

\section{X--ray or opt--UV black--body as supernova shock breakout?}

Since we have discarded the possibility of a single 
(or a multi--color) \bb joining the opt--UV and 
the X--ray emission, let us discuss the case of
a supernova shock breakout associated with either
the opt--UV emission or the X--ray \bbp.

Assume first that the association is with the X--ray \bbp.  
Detailed modelling of the shock breakout (Li 2006) flashes in SN
explosions predicts a temperature of 1.8 keV (0.2 keV), a duration of
$\sim$20 s and an emission in the X--ray band amounting to a
radiated energy of $\sim 10^{47}$ erg for a typical hypernova,
and $\sim 10^{45}$ erg for ``normal" SNe Ibc. 
These
values are inconsistent with the temperature, duration and energetics
of the observed X--ray \bb component of GRB 060218. 
In addition
to this, the radii derived from the \bb fit to the X--ray data (C06)
are increasing in time, but at a rate corresponding to a very small
velocity: $\sim$3000 km s$^{-1}$, in contrast with the (decreasing)
velocities derived by the optical spectroscopy by Pian et al. (2006),
which shows velocities of 20,000 km s$^{-1}$ at one day from the
trigger. 

Consider now the association with the opt--UV.
In this case one can assume that the flux in this band
belongs to a \bb peaking at (or not too above) the largest 
observed frequency (to limit the implied energetics),
but in this case the corresponding \bb radius is around
$10^{15}$ cm.
Then, if the supernova exploded around the same time
of the GRB, the implied velocity of the ejecta exceeds $c$.

In summary:
a single \bb joining the opt--UV and the X--ray fluxes is too energetic;
the \bb emission expected from a SN shock breakout cannot
be associated with the \bb observed in the X--rays (too small
inferred velocities of the ejecta), nor with a \bb 
peaking in the UV (too large ejecta velocities).

We are then forced to explore alternative possibilities to explain the
SED of this burst.  
One possibility is that the entire SED (except the X--ray 
\bb component) is produced by 
the synchrotron process, which can account for the
optical--UV very hard spectrum if it self--absorbs
at or above the UV band.
This model will be discussed in the next section.

\section{A synchrotron self--Compton model} 

Assume that the overall opt--UV to X--ray SED, excluding the
X--ray black--body, belongs to the same synchrotron spectrum.
For simplicity assume that this radiation is produced in 
a jet, with cross sectional radius $R$, 
width $\Delta R^\prime$ (as measured in the comoving frame),
semi--aperture angle $\psi$, embedded in a tangled magnetic 
field $B$, moving with a bulk Lorentz factor $\Gamma$.
We assume that the radiation we observe
is produced at a fixed distance form the jet apex.
In other words, the conversion of bulk kinetic into
random energy occurs at the same location along the jet,
for the entire duration of the burst.
The viewing angle is supposed to be smaller than $\psi$.
Assume also that the emitting particles are
distributed in energy according to a simple power law
$N(\gamma)=K\gamma^{-p}$ between $\gamma_{\rm min}$ 
and $\gamma_{\rm max}$.
We require that the synchrotron spectrum self--absorbs
at a frequency $\nu_{\rm a}$ close to $10^{15}$ Hz.
If $\gamma_{\rm min}$ is small (around unity), then
the self absorbed spectrum should be $\propto \nu^{5/2}$.
Note that the opt--UV spectrum shown in Fig. 1 
{\it assumed} a spectral slope $\propto \nu^2$ to derive
the extinction in the host frame (C06). 
A spectrum $\propto \nu^{5/2}$ requires a small increase
in the host extinction: $E(B-V)_{\rm host}$ increases 
from 0.2 to 0.3. 

This is shown in Fig. 2, together with the results of the synchrotron
self--Compton model (dashed lines), at $\sim$ 2000 s, $\sim 7000$ s
and at a time between $10^4$ and $10^5$ s.

For the 2000 s spectrum, we have used $R=7\times 10^{11}$ cm, $\Delta
R^\prime=10^{11}$ cm, $\Gamma=5$, a semi--aperture angle of the jet
$\psi=0.2$, $B=3\times 10^5$ G, and $p=2.3$, between $\gamma_{\rm
min}=1$ and $\gamma_{\max}=360$.  
We assumed a Doppler factor $\delta\sim 2\Gamma$, appropriate for 
on--axis observers (but the results are nearly the same for observers 
within the jet opening angle).  
For the comoving intrinsic luminosity we set
$L^\prime=6.5 \times 10^{42}$ erg s$^{-1}$.  
To obtain the
isotropically equivalent luminosity we used $L =L^\prime
\delta^2/(1-\cos\psi)$, while the monochromatic luminosities have been
boosted by $L(\nu) =L^\prime (\nu^\prime) \delta/(1-\cos\psi)$.  The
self--Compton luminosity is a factor $\sim 30$ less then the
synchrotron one.

For the 7000 s spectrum we 
can describe the spectrum by changing only 
the slope of the electron distribution from 2.3 to 4.2,
and mantaining constant all other parameters.
The choice of $p$ is not free, since it is dictated
by the measured slope of the X--ray continuum.
This ``drastic" change of $p$ on such a short timescale
may appear odd, but it is not unprecedented: in fact blazars
show such a behavior quite often (see e.g. Mkn 501
in Sambruna et al. 2000).

For the late SED (at $10^4$--$10^5$ s), we again changed only the
electron distribution function, introducing a break at 
$\gamma_{\rm break}=7.4$.  
Below this break $p=2$, above it $p=15$, a value so
large to mimic an exponential roll--off.  
The observed radiation in
the X--rays corresponds to the first order self--Compton scattering.
Within this scheme, since the opt--UV flux remains almost the same,
and is described by an absorbed spectrum, we have that $R^2 B^{-1/2}$
must be the same as before.  Therefore this radiation {\it is not
afterglow, but late prompt emission} (as also proposed by Fan et
al. 2006).  Since the slope of the electron distribution is much
steeper, what we see in the X--rays at such late times is the first
order self--Compton spectrum. 
This explains why we observe an uncommonly soft X-ray spectrum at 
times $>10^{4}$ s.

Using the chosen parameters, the value of the corresponding
Poynting flux $L_B=\pi R^2\Gamma^2 cB^2/(8\pi)=4\times 10^{45}$
erg s$^{-1}$. 
If this corresponds to a conserved quantity, then the value of $B$ 
at a distance of $5\times 10^5$ cm (where $R=10^5$ cm)
from the central power--house is $B_0 = 2\times 10^{12}$ G.
The value of $L_B$ is much greater than the kinetic power
carried by the protons associated with the emitting
electrons (assuming one proton per electron), but there is
the possibility that only a fraction of leptons are accelerated.
In this case the kinetic energy of protons would increase.

\begin{figure}
\vskip -0.5 true cm
\hskip -0.7 true cm \psfig{file=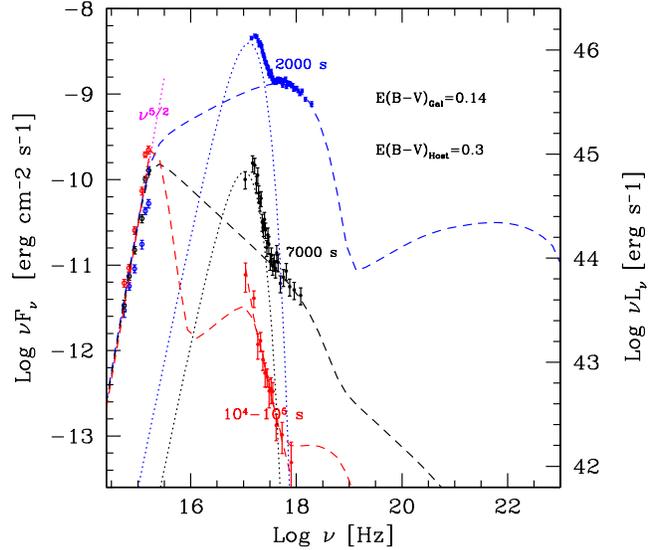,height=8.7cm,width=9cm}
\vskip -0.5 true cm
\caption{ 
The SED of GRB 060218 at different times, as in Fig. 1, but
with the optical UV points de--reddened with $E(B-V)_{\rm host}=0.3$
instead of 0.2 This produces an opt--UV spectrum $\propto \nu^{5/2}$.
We also show the SSC model, discussed in the text, 
for the 3 SEDs for which we have simultaneous UVOT, XRT data 
(i.e. at 2000, 7000 and $\sim 10^4$--$10^5$ seconds after trigger).  
}
\label{fig2}
\end{figure}

\begin{figure}
\vskip -0.5 true cm
\hskip -0.7 true cm \psfig{file=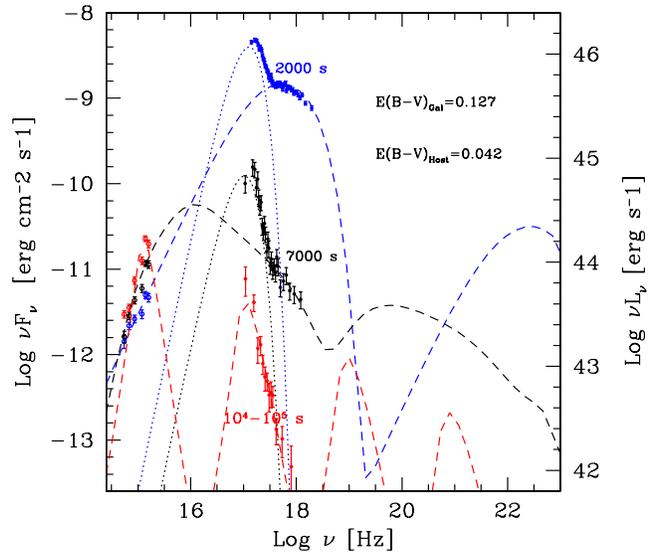,height=8.7cm,width=9cm}
\vskip -0.5 true cm
\caption{ 
As in Fig. 2, but
with the optical UV points de-reddened with $E(B-V)_{\rm Gal}=0.127$
and $E(B-V)_{\rm host}=0.042$.
We show the SSC model, derived in this case,
as discussed in the text.  
}
\label{fig3}
\end{figure}


The self absorbed synchrotron luminosity
$L^\prime (\nu^\prime)/{\nu^\prime}^{5/2} \propto R^2 B^{-1/2}$, 
while for the optically thin part 
$L^\prime (\nu^\prime){\nu^\prime}^{\alpha}
\propto R^2 \Delta R^\prime K B^{1+\alpha}$ where 
$\alpha=(p-1)/2$ is the energy spectral index.
Including $\delta$, we then have 5 unknowns 
($\delta, R, \Delta R^\prime, B, K$)
and only two observables (the optically thick and thin
part of the spectrum).  
The cut--off energy fixes directly $\gamma_{\rm max}$ once 
$B$ and $\delta$ are given, and the observed slope directly fixes $p$.

The found solution therefore is not unique, 
but we were guided in the
choice of the parameters by some additional constraints:
i) the Comptonization 
$y\sim \sigma_{\rm T} K \Delta R^\prime \langle \gamma^2\rangle$ 
parameter should not be greater than unity,
to not produce too much self--Compton emission;
ii) the needed self--absorption frequency is rather
large, and requires a large value of the magnetic field, 
suggesting a small size of the source.
However, a lower limit to the size of the emission region can be
obtained by requiring that it is transparent to 
Thomson scattering;
iii) in strong magnetic fields, the radiative cooling times 
are much shorter than the dynamical times, and all electrons
can cool down to $\gamma_{\rm min}=1$.
These additional constraints, however, are not enough to single out
a unique solution, and the models shown in Fig. 2 should be
considered as illustrative examples only.
This SSC model, on the other hand, can explain in a simple
way why the optical UV flux changes much less than the X--ray flux,
since it is due to the self--absorbed flux, much less sensitive
to changes in the electron distribution.

Another concern is the adopted value of the optical extinction, which
is uncertain.  Sollerman et al. (2006) proposed a significant lower
value: $E(B-V)_{Gal}=0.127$ and $E(B-V)_{host}=0.042$.  In this case
the de--reddened opt--UV flux does not describe a $\nu^{5/2}$ law,
being approximately intermediate between $\nu^{1/3}$ and $\nu^2$.  In
Fig. 3 we show our SSC model in this case: the parameters are similar
to the ones used for Fig. 2, the differences being in the chosen
values of $\gamma_{\rm min}$ (now equal to 100 for the 2000 s
spectrum, 35 for the 7000 s spectrum and 7.1 for the $10^4$--$10^5$ s
spectrum).  We had to change the size and the magnetic field value
from $R=7\times 10^{11}$ cm and $B=3\times 10^5$ G for the 2000 s
spectrum to $R=2.5\times 10^{11}$ cm and $B=10^5$ G for the other two
spectra.  The other parameters are the same as for the model in
Fig. 2.

Within this SSC scenario the ``afterglow" of GRB 060218 is instead
 late prompt emission.  The quasi--exponential decline in the X--ray
 light curve after roughly 3000 seconds is produced by the synchrotron
 tail going out from the observed 0.2-10 keV X--ray energy window,
 while the subsequent power law decay is produced by the first order
 self Compton emission entering (and remaining) in this energy band.
 Since the electron distribution becomes steeper and steeper in time,
 this may explain why the temporal decay of the X--ray ``afterglow"
 seems normal, while the spectral slope is much steeper than in the
 majority of GRBs: being prompt emission, no closure relation can be
 applied to the light curve and spectrum of GRB 060218.

\section{The X--ray Black--body}

\begin{figure}
\vskip -0.5 true cm
\hskip -0.7 true cm \psfig{file=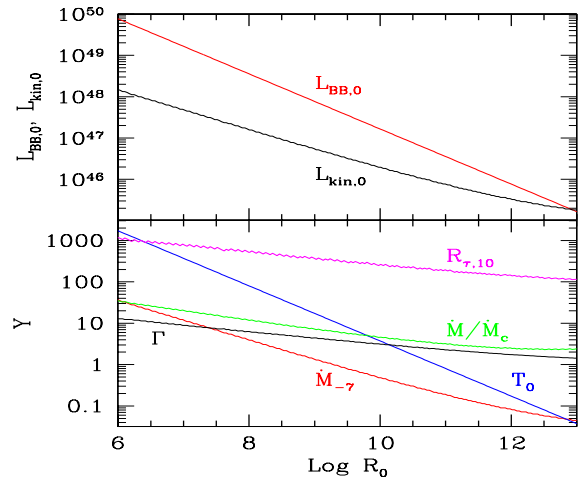,height=7.5cm,width=10cm}
\vskip -0.5 true cm
\caption{
Top panel: the initial \bb and kinetic powers
as a function of the dissipation radius $R_0$,
calculated assuming fixed values of the observed \bb temperature 
(equal to 0.18 keV) and luminosity 
(equal to $8\times 10^{45}$ erg s$^{-1}$).
We assumed an initial velocity equal to $\beta=1/\sqrt{3}$
corresponding to $\Gamma_0=1.225$.
Bottom panel: the mass outflowing rate $\dot M$, the ratio
$\dot M/\dot M_{\rm c}$, the final bulk Lorentz factor $\Gamma$,
the transparency radius $R_\tau$ as a function of the
dissipation radius $R_0$.
Note that $\dot M_{\rm c}$ is defined
by setting $R_\tau=R_{\rm acc}$.
For $\dot M$ above $\dot M_{\rm c}$ the material
completes its acceleration before it becomes
transparent, while the opposite occurs for 
$\dot M <\dot M_{\rm c}$. 
}
\label{fig4}
\end{figure}


Here we try to explain
the nature of the soft \bb ($kT\sim$0.1--0.18 keV) component
observed in the X--ray band up to 7000 s. 
It has been proposed that
this soft X--ray thermal component corresponds to the shock breakout
in a dense wind--like circumburst environment\footnote{
Note that if instead the shock breakout happens in a normal
stellar envelope the required progenitor's radius is $\sim 10^{12}$ cm
which is much larger than what expected for an H--He envelope--stripped
progenitor as suggested from spectroscopy (Pian et al. 2006) } 
(C06) although theoretical modelling seems to rule out this 
possibility (Li 2006). 
Alternative explanations (Fan, Piran \& Xu 2006) invoke the
thermal emission from a hot cocoon (e.g. Ramirez--Ruiz, Celotti \&
Rees 2002) surrounding the jet.  
It was also proposed that the non--thermal prompt component 
observed in GRB 060218 might result from the bulk Comptonization 
of the soft X--ray thermal photons by a mildly relativistic jet
(Wang, Li, Waxman \& Meszaros 2006).

Here we investigate if the observed soft X--ray \bb
component might be interpreted within the classical GRB fireball
model. 
We consider two possible scenarios,
namely the \bb we see in the X--rays is the leftover from the
initial acceleration, or it is the result of some dissipation 
(Rees \& Meszaros 2005; Pe\'er, Meszaros \& Rees, 2006;
Thompson 2006) occurring
later, inside the star or in the vicinity of its surface.  
In the
former case we expect that the {\it initial} \bb has a large
temperature and luminosity, while in the latter the \bb ``degrades"
much less, and the observed temperature and luminosity are not vastly
different from their initial values.

In the case of an adiabatic expansion, due to the conversion
of internal energy into bulk motion, the dynamics is
controlled by four parameters:
the initial radius $R_0$, the initial luminosity $L_0$ of the
radiation assumed to be responsible for the expansion,
the initial bulk Lorentz factor $\Gamma_0$ and the outflow
mass rate $\dot M$, assumed to be constant during the expansion.
The initial kinetic power of the material is
$\dot E_{\rm kin,0}=\Gamma_0 \dot M c^2$.
We then follow the usual prescription of an adiabatic expanding
fireball.

According to this scenario, the temperature of the
internal radiation
is observed to be constant as long as we are in the acceleration phase,
while it decreases as $R^{-2/3}$ between $R_{\rm acc}=\Gamma R_0/\Gamma_0$ 
(the radius where the acceleration ends) and $R_\tau$, where
the fireball becomes transparent (if $R_\tau>R_{\rm acc}$).
The value of $R_\tau$ is (Daigne \& Mochkovitch 2002; Meszaros 2006)
\begin{equation}
R_\tau =  {\sigma_{\rm T}  Y \dot M_{\rm f, iso} 
\over 8\pi m_{\rm p} c \Gamma^2 } =
{\sigma_{\rm T}  Y \dot E_{\rm k, iso} 
\over 8\pi m_{\rm p} c^3 \Gamma^3 } = 
2.93\times 10^{13} {\dot E_{\rm 47, iso} \over \Gamma^3}\,  {\rm cm}
\end{equation}
where $Y=0.5$ is the number of electrons per barion
and $\dot E_{\rm 47, iso}$ is the kinetic power of the fireball
(i.e. $\dot E_{\rm k,iso} =\Gamma \dot M c^2$) 
in units of $10^{47}$ erg s$^{-1}$.
At $R_\tau$ the observed \bb temperature $T_{\rm ph}$ is
\begin{equation}
T_{\rm ph} \, = T_0 \left( { R_{\rm acc}\over R_\tau}\right)^{2/3}
\, \propto \, \left( {R_0 \Gamma^4 \over 
\Gamma_0\dot E_{\rm kin}}\right)^{2/3}
\end{equation}
where $T_0$ is the initial temperature.  The photon number is
conserved, hence
\begin{equation}
{L_0 \over T_0} \, \sim \, 4\pi {R_0^2\over \Gamma_0^2}
 \sigma T_0^3  =  
{L_{\rm ph} \over T_{\rm ph}}  \to \, L_{\rm ph}  = 
4 \pi {R_0^2\over \Gamma_0^2} \sigma T_{\rm ph}^4 
\left( { R_\tau \over R_{\rm acc}}\right)^2
\end{equation}
For any assumed value of $R_0/\Gamma_0$, from
Eq. 2 and Eq. 3 one obtains $T_0$ and $L_0$ as
a function of the observables $T_{\rm ph}$ and $L_{\rm ph}$.
If we fix only $\Gamma_0$, then $T_0$ and $L_0$ are
functions of $R_0$. 

Furthermore, if the power of the fireball is
the sum of its initial kinetic plus radiation
powers, we can derive $\dot E_{\rm kin}$
and by knowing $T_0$ and $T_{\rm ph}$ we can 
derive the final value of $\Gamma$ as a function
of $R_0$, again for a specific value of $\Gamma_0$.
Finally, knowing $E_{\rm kin}$ and $\Gamma$, we can 
derive $\dot M$.

All these quantities are shown in Fig. 4 (as a function
of $R_0$) for the specific choice of $\Gamma_0=1.225$,
corresponding to an initial bulk velocity equal to
the sound speed of a relativistic plasma. 
We can see that if we assume that the observed X--ray \bb
is the leftover of the initial radiation, injected at $R_0\sim 10^6$
cm, then the required initial luminosity is very large, and
since this power should last for an unusually long time (for this
burst), the isotropically equivalent total energetics becomes huge.
Consider instead values of $R_0$ around $10^{11}$
cm: in this case the \bb is the result of some later
dissipation, either in the fireball itself or
resulting from the interaction of the fireball with
the material forming the cocoon or at the surface of the
star.

Since the dissipation occurs in this case much closer to the
transparency radius, the \bb does not ``degrade" much,
being consistent with what we see without implying 
very large initial luminosities.
For $R_0\sim 10^{11}$ we have $\Gamma\sim 2$,
$\dot M\sim 10^{-8}~ M_\odot$ s$^{-1}$, $T_0$ around 1 keV
and $R_\tau \sim 10^{12}$ cm.

\section{Summary and conclusions}

We have studied the broad band opt--UV to X--ray SED at three epochs
which characterise different flux evolution phases observed (C06) both
in the X--ray and opt--UV light curves up to $10^{5}$ s. 
The spectrum
shows a soft ($kT\sim$ 0.1--0.18 keV) X--ray \bb component
(coupled with a typical non--thermal component) together with a
Rayleigh--Jeans tail in the opt--UV band. 
Both these components are
almost steady in flux and slope, while the X--ray \bb is undetected at
late times (10$^4$--10$^5$ s).
They have been interpreted (e.g. C06) as the shock breakout of the 
accompanying (nearly simultaneous) SN2006aj. 
However, if the X--ray and opt--UV emission belong to the
same \bb emission (which we also exclude by direct spectral
fitting) its energetics is even larger than the total kinetic energy of
the SN ejecta as derived from its late time spectroscopy (e.g. Mazzali
et al. 2006; Pian et al. 2006), 
and cannot be cured by relativistic
beaming, since the SN shock is not relativistic.
 Alternatively, only the X--ray or the
opt--UV thermal component might be the shock breakout signal. The
former possibility seems to be excluded both by detailed simulations
(Li 2006) and because the SN expansion velocity, derived from the
X--ray \bb fits, is too small compared to that derived from optical
spectroscopy at late times (Pian et al. 2006). 
On the other hand, a
shock breakout in the opt--UV band would imply a too large SN expansion
velocity to be consistent with the observed opt--UV early spectrum.

If the SN shock breakout went undetected both in the opt--UV and in the
X--ray band we are left with two puzzling quests: (a) which is the
nature of the observed early time opt--UV emission
(i.e. before the SN radioactive decay emission sets in at 10$^{5}$ s)  
and (b) what is the origin of the X--ray thermal component?  

We explored the possibility that the SED is produced by synchrotron
self--Compton emission. 
A simple synchrotron fit to these data is satisfactory,  
with the opt--UV data corresponding to the
self--absorbed synchrotron spectrum.
Although the choice of the input parameter is not unique, the spectra 
at different times 
can be modelled by changing the slope of the electron distribution.  
A remarkable result of these fits is that also
the late time (10$^{4}$--10$^{5}$ s) opt--UV emission should be
described by the same model (i.e. with the same parameters of the
prompt, but with steeper still slopes of the electron distribution).
The X--ray flux is in this case first order self--Compton emission and
naturally accounts for the unusually soft spectrum observed (which is
hardly reconcilable with the normal flux time decay within the
context of the standard afterglow model -- Fan, Piran \& Xu 2006). 

Finally, we considered the possibility that the \bb observed in the
X--ray band is a separate component.  
It is consistent to be
photospheric emission from the cocoon or some stellar material
energized by the GRB jet. 
The origin of the \bb photons should correspond to dissipation 
occurring not much below the photospheric radius of this material.

\section*{Acknowledgements}
We thank S. Campana for useful discussion and for having provided us
the updated UVOT light curves\\ (see also
http://www.brera.inaf.it/utenti/campana/060218/060218.html).

\end{document}